# CCAT-NET: A NOVEL TRANSFORMER BASED SEMI-SUPERVISED FRAMEWORK FOR COVID-19 LUNG LESION SEGMENTATION

Mingyang Liu[1,3]   Li Xiao[1,3,4*]   Huiqin Jiang[2*]   Qing He[1,3]

[1] Henan Institutes of Advanced Technology and Research Center of Digital Medical Image Technique, Zhengzhou University, Zhengzhou 450052, P.R. China
[2] School of Information Engineering and Research Center of Digital Medical Image Technique Zhengzhou University
[3] Key Lab of Intelligent Information Processing of Chinese Academy of Sciences (CAS), Institute of Computing Technology, CAS, Beijing 100190, China
[4] Ningbo Huamei Hospital, University of the Chinese Academy of Sciences, Ningbo, China

## ABSTRACT

The spread of the novel coronavirus disease 2019 (COVID-19) has claimed millions of lives. Automatic segmentation of lesions from CT images can assist doctors with screening, treatment, and monitoring. However, accurate segmentation of lesions from CT images can be very challenging due to data and model limitations. Recently, Transformer-based networks have attracted a lot of attention in the area of computer vision, as Transformer outperforms CNN at a bunch of tasks. In this work, we propose a novel network structure that combines CNN and Transformer for the segmentation of COVID-19 lesions. We further propose an efficient semi-supervised learning framework to address the shortage of labeled data. Extensive experiments showed that our proposed network outperforms most existing networks and the semi-supervised learning framework can outperform the base network by 3.0% and 8.2% in terms of Dice coefficient and sensitivity.

## 1. INTRODUCTION

Since December 2019, the spread of the novel coronavirus disease 2019 (COVID-19)[1] has put the world in a huge existential crisis. By the end of November 2020, there had been close to 63 million reported cases of COVID-19 globally and over 1.4 million deaths[2]. As a complement to the RT-PCR test, the computed tomography (CT) has proven its effectiveness in current diagnostics, including follow-up assessments and evaluation of disease evolution[3]. Doctors can observe typical signs of lesion on CT scans such as ground-glass opacity (GGO), which can help them to diagnose and treat patients. However, segmentation of COVID-19 lesion in CT scans remains a challenging task due to issues such as the highly variable texture, size and location of the lesion in CT scans.

Convolutional Neural Network (CNN) is widely used for the task of image segmentation. There has been a lot of work to apply CNN to medical images' segmentation such as U-net[4] and Unet++[5]. Several studies have also designed network structures specifically for the segmentation of COVID-19 lesions such as Inf-Net[6] and MiniSeg[7]. Recently, Transformer-based networks lead a new trend in the computer vision field. However, they have not been applied much to medical images because they need to be trained on large-scale datasets to achieve excellent performance. Since manually labeling infected regions is a tedious and laborious task, COVID-19 lesions images with high-quality labels are often very scarce making it difficult for the Transformer to deliver its performance.

To alleviate the network's dependence on labeled data and enhance its performance, we propose CCAT-net, a novel network structure that combines CNN and Transformer. CCAT-net extracts local features by CNN, global features by Transformer and fuses high-resolution information with low-resolution information by skip connections. It has both the fast convergence of CNN and the powerful representation capability of Transformer. We further incorporate it with an efficient semi-supervised learning framework to make full use of unlabeled data.

Our contributions in this work are threefold:(1) We introduce **a novel CCAT-net** that realizes automatic and efficient segmentation of COVID-19 lesions by combining the respective strengths of CNN and Transformer. (2) We propose **an efficient semi-supervised learning framework** that unifies dominant ideas such as consistency regularization and Pseudo-labeling to alleviate the shortage of labeled data. (3) Extensive experiments showed that our proposed network **outperforms** most existing networks and the semi-supervised learning framework can further improve the learning capability of our network.

---
* Correspondence: Li Xiao(andrew.lxiao@gmail.com, xiaoli@ict.ac.cn), Huiqin Jiang(iehqjiang@zzu.edu.cn)

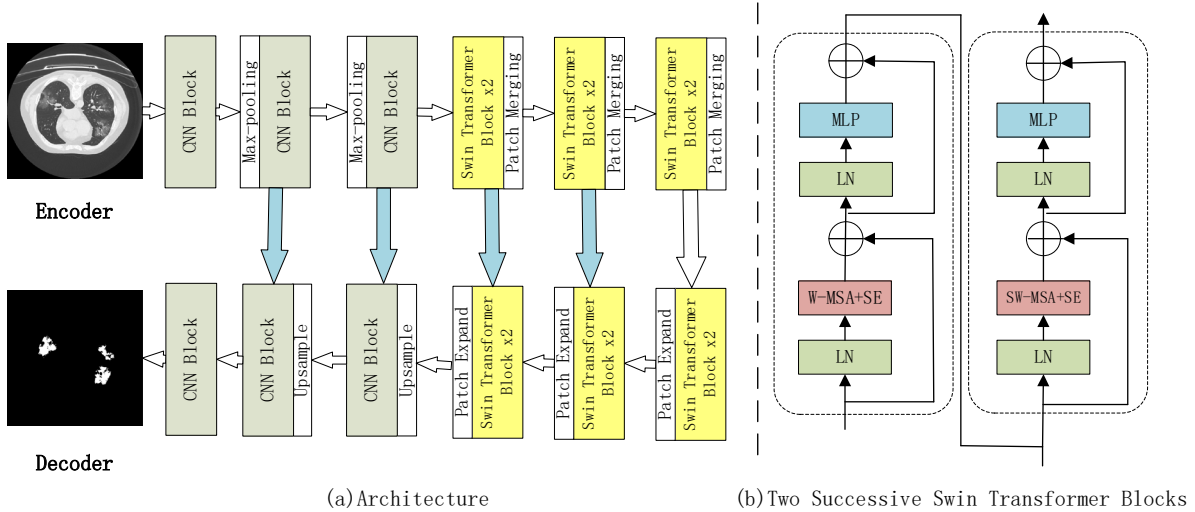

**Fig. 1.** (a) The proposed network architecture for Covid-19 segmentation; (b) Two Successive Swin Transformer Blocks.

## 2. METHOD

### 2.1. Overview of network architecture

The base architecture of the proposed CCAT-net is presented in Figure 1. (a). The baseline of CCAT-net consists of encoder, decoder, and skip connections. The encoder and decoder mainly consist of Convolution Neural Networks(CNN) blocks and Swin Transformer blocks. CNN and Transformer are complementary since CNN is better at extracting local features, while Transformer is more effective in extracting long-range dependencies. Given their respective characteristics, we adopt the CNN block to extract shallow low-level local information and the Swin Tranforer block to extract deep high-level semantic information.

**CNN block**  A CNN block consists of the repeated application of two 3x3 convolutions, each followed by Batch Normalization and an activation function LeakyRelu.

**Swin Transformer block**  A Swin Transformer block consists of a shifted window based MSA module, followed by a 2-layer MLP with GELU non-linearity in between. A LayerNorm (LN) layer is applied before each MSA module and each MLP, and a residual connection is applied after each module[8]. The architecture of two successive Swin Transformer Blocks is presented in Figure 1. (b). In particular, we add the SE module[9] after W-MSA and SW-MSA to filter redundant information in the channels[10].

**Patch Merging layer** concatenates the features of each group of 2×2 neighboring patches and applies a linear layer on concatenated features. It is a downsampling operation in ViT. **Patch Expand layer** is just the opposite operation of Patch Merging layer.

The input image first passes through three CNN blocks to extract shallow low-level local information. The second and third CNN blocks contain a max-pooling layer at the beginning to perform downsampling. Then the feature map is transformed into patch embedding that is suitable for Transformer's input. After passing two successive Swin Transformer Blocks and Patch Merging three times, the model completes the feature extraction of the encoder part. The decoder and encoder are almost perfectly symmetrical. The downsampling part is replaced by the corresponding upsampling. The corresponding blocks of the encoder and decoder are connected via skip connections to fuse the multi-scale features from the encoder with the decoder's upsampled features.

### 2.2. Semi-supervised learning framework

We further incorporate the base network of CCAT-net with a semi-supervised framework to improve its performance. The overall architecture of the proposed semi-supervised learning framework is presented in Figure 2.

The framework adopts the **mean teacher structure**, the teacher model uses the exponential moving average (EMA) weights of the student model[11] to improve the target quality. The training objective contains four loss items: supervised loss $L_s$, consistency loss $L_c$, mixup loss $L_m$ and fix loss $L_f$. $L_c$ mines the intrinsic information of unlabeled images by encouraging the model to give consistent outputs for perturbed versions of the same image. $L_m$ encourages models to make better generalizations about unseen data by mixing labeled and unlabeled data. $L_f$ enables the model to learn more abstract invariants by strongly augmenting the data more abstractly. Next, we describe them in detail.

**Supervised loss** is used for gradient minimization between the student model's prediction of **labeled images** $p_l$ and the labels y. We define $L_s$ as a combination of the cross entropy loss $L_{ce}$ and the tversky loss $L_{tversky}$:

$$L_s = \alpha_1 L_{ce}(p_l, y) + \alpha_2 L_{tversky}(p_l, y) \quad (1)$$

In our experiments, we set $\alpha_1$=0.5 and $\alpha_2$=0.5.

**Consistency loss** uses perturbation invariance to enrich the learned representation of the model. We apply K=2 (random dropout and noise) different kinds of **weak data augmentation** to **unlabeled images** and use $L_c$ to encourage the model to generate similar predictions. Denoting the student model's predictions of unlabeled images as $p_u$, the teacher model's predictions of augmented images as $p_{w1}$ and $p_{w2}$. So we define **$L_c$** as the mean square error (MSE) between predictions of the student model and the predictions of the teacher model.

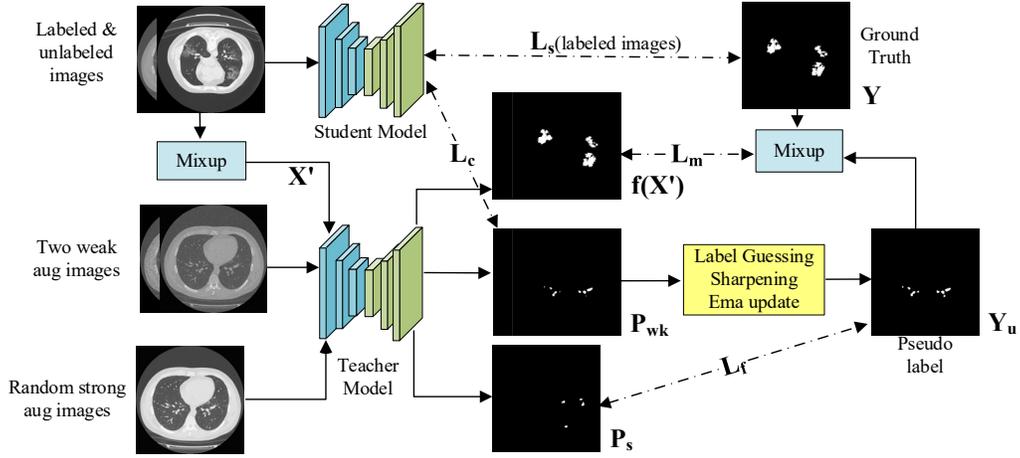

**Fig. 2.** The proposed semi-supervised learning framework for Covid-19 segmentation.

$$L_c = \frac{1}{K \cdot N} \sum_{k=1}^{K} \sum_{i=1}^{N} (p_u^i - p_{wk}^i)^2 \quad (2)$$

Where i and N are the index and the total number of unlabeled images, K is the number of data augmentation.

**Label guessing**: For unlabeled images, we use the model's predictions to guess labels for them[12]. *This guess is later used as **pseudo labels** in the $L_m$ and $L_f$.* To do first, we compute the average of the model's prediction across unlabeled images and their perturbed versions.

$$\overline{p} = \frac{1}{K+1} \left( \sum_{k=1}^{K} p_{wk} + p_u \right) \quad (3)$$

**Sharpening & EMA update**: For reducing the entropy of the label distribution, we perform a sharpening operation on the average prediction by adjusting the "temperature" of the categorical distribution. We update the sharpen result with EMA to improve its quality. The operations are defined as:

$$Y_u^t = Sharpen(\overline{p}, T) = \overline{p}^{\frac{1}{T}} / \sum_{i=1}^{n} \overline{p}_i^{\frac{1}{T}} \quad (4)$$

$$Y_u^t = \alpha Y_u^{t-1} + (1-\alpha) Y_u^t \quad (5)$$

Where t is the number of training iteration and α is a smoothing coefficient hyperparameter, we set T=0.5 and α=0.9 in our experiments.

**Mixup** mixes **labeled and unlabeled data** to make it harder for the model to remember the training data and therefore hopefully make it better at generalizing to unseen data. Denoting the labeled images as X, the GT of them as Y, the unlabeled images as $X_u$, $Y_u$ as their pseudo label, the teacher model as f(), Mixup and **$L_m$** are defined as:

$$X_1 = Concat(X, X_u), \ Y_1 = Concat(Y, Y_u) \quad (6)$$
$$X_2 = Shuffle(X_1), \ Y_2 = Shuffle(Y_1) \quad (7)$$
$$\lambda = Beta(\alpha, \alpha) \quad (8)$$
$$\lambda' = \max(\lambda, 1-\lambda) \quad (9)$$
$$X' = \lambda' X_1 + (1-\lambda') X_2 \quad (10)$$
$$L_m = \lambda' L_{ce}(f(X'), Y_1) + (1-\lambda') L_{ce}(f(X'), Y_2) \quad (11)$$

Where λ is sampled from the beta distribution and α is a hyperparameter, we set α=0.75 in our experiments.

**Fix loss:** We apply random **strong data augmentation** to **unlabeled images** to enable the model to learn more abstract invariants. In detail, we selected eight strong data augmentations suitable for the segmentation task from the python library imgaug and used one of them at random each time. Denoting the model's prediction of them as $P_s$, the $Y_u$ as their pseudo label, the **$L_f$** is defined as:

$$L_f = L_{mse}(P_s, Y_u) \quad (12)$$

Finally, we define the **total loss function L** as:

$$L = L_s + \alpha L_c + \beta L_m + \gamma L_f \quad (13)$$

α, β and γ are coefficients represent importance of the different losses. In our experiments, we set α=0.3, β=0.4, and γ=0.3.

## 3. EXPERIMENTS

### 3.1. Datasets and Evaluation Metrics

**Datasets** The dataset COVID-19-P1110 we use is provided by MiniSeg's authors. It contains 785 2D CT images and the corresponding expert annotations of 50 patients. We use random rotating and flipping for data augmentation. We perform 5-fold cross-validation to avoid statistically significant differences in performance evaluation[7]. The unlabeled CT images are sampled from the same dataset Mosmeddata[13]. When we use the semi-supervised learning framework, the ratio of labeled data to unlabeled data is 1:1 in each fold.

**Evaluation Metrics** We adopt Dice similarity coefficient

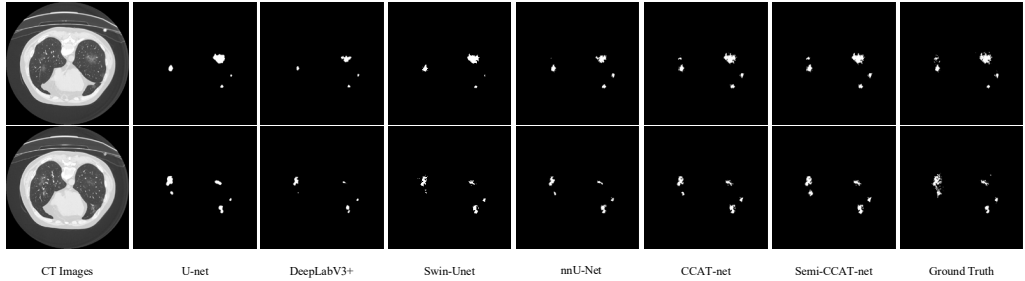

**Fig. 3.** Visual comparison of lung lesion segmentation results for partial methods.

(DSC), Hausdorff distance (HD), sensitivity (SEN) and specificity (SPC) as evaluation metrics.

### 3.2. Implementation details

The framework we use is Pytorch. We use SGD optimizer for training with the momentum of 0.9 and the weight decay of 1e-4. The learning rate policy we adopt is warmup and poly, where the initial learning rate is 0.01. As the Transformer is more difficult to converge than CNN, we train 200 epochs on the training set with a batch size of 5 for CCAT-net and Swin-Unet. For other CNN-based networks, we train 80 epochs. Since Swin-Unet has to load pre-trained weights, the resolution of input image is 224×224 and the resolution of the input image of other models is 512×512. Apart from the above differences, other settings we use are the same for a fair comparison. For semi-supervised learning, we fine-tune 100 epochs for each fold on the weights that performed best on validation set in previous fully supervised learning.

### 3.3. Comparison with other models

We compare CCAT-net to other state-of-art methods. Quantitative results are shown in Table 1. The segmentation results for partial methods are shown in Figure 3.

**Table 1**. Comparisons among different methods.

| Method | DSC | HD | SEN | SPC |
|---|---|---|---|---|
| U-net[4] | 0.5279 | 72.45 | 0.5471 | 0.9018 |
| Unet++[5] | 0.5873 | 80.94 | 0.6521 | 0.9471 |
| Inf-net[6] | 0.5723 | 78.60 | 0.6318 | 0.9457 |
| PSPNet[14] | 0.5878 | 68.22 | 0.6695 | 0.9568 |
| FPN[15] | 0.6072 | **62.16** | 0.6644 | 0.9546 |
| DeepLabV3[16] | 0.5828 | 69.05 | 0.7223 | 0.9635 |
| DeepLabV3+[17] | 0.6149 | 62.91 | 0.6639 | 0.9566 |
| Swin-Unet[18] | 0.5766 | 87.55 | 0.6615 | 0.9604 |
| nnU-Net[19] | 0.6383 | 63.61 | 0.6749 | 0.9511 |
| CCAT-net(ours) | 0.6316 | 81.68 | 0.7022 | 0.9758 |
| CCAT-net(with semi) | **0.6506** | 78.51 | **0.7600** | **0.9766** |

The performance of CCAT-net has improved considerably compared to most previous SOTA methods. By making full use of the unlabeled data, our semi-supervised learning framework improves the network performance even further.

As our network is a combination of CNN and Transformer, we conducted comparative experiments using U-net and Swin-Unet without loading the pre-trained weights. All the three networks are trained for 200 epochs.

**Table 2**. Further comparative experiments.

| Method | DSC | HD | SEN | SPC |
|---|---|---|---|---|
| U-net(depth=6) | 0.5950 | 76.21 | 0.6836 | 0.9438 |
| Swin-Unet($512^2$) | 0.0804 | 186.12 | 0.1675 | 0.9397 |
| CCAT-net | 0.6316 | 81.68 | 0.7022 | 0.9766 |

As shown in Table 2, CCAT-net converges well and outperforms **fully CNN-based** U-net, with an increase of DSC as 6.0%, SEN as 2.3% and SPC as 3.5%. It is also worth noting that the **fully Transformer-based** Swin-unet (without pre-trained weights) cannot even converge and does not have an acceptable prediction.

### 3.4. Ablation study

To demonstrate the effectiveness of each loss item in our semi-supervised learning framework, we performed ablation experiments. The experimental results in Table 3 show that all the four loss items continuously improve the performances.

**Table 3**. Ablation studies of each loss item.

| $L_s$ | $L_c$ | $L_m$ | $L_f$ | DSC | HD | SEN | SPC |
|---|---|---|---|---|---|---|---|
| √ | | | | 0.6316 | 81.68 | 0.7022 | 0.9758 |
| √ | √ | | | 0.6388 | 82.86 | **0.7842** | 0.9749 |
| √ | √ | √ | | 0.6455 | 80.03 | 0.7656 | 0.9703 |
| √ | √ | √ | √ | **0.6506** | **78.51** | 0.7600 | **0.9766** |

## 4. CONCLUSIONS

In this paper, we propose a novel CCAT-net that combines CNN and Transformer for the segmentation of COVID-19 lesions. Extensive experiments showed that our proposed network outperforms most previous SOTA methods based on CNN. Different from previous Transformer-based networks, our network does not need to load pre-trained weights to achieve excellent results on small datasets, which gives a new way of thinking about the application of Transformer to small datasets. To address the shortage of high-quality labeled data, we propose an efficient semi-supervised learning framework that can make full use of unlabeled data to outperform the base network by 3.0% and 8.2% in terms of DSC and SEN.

## 5. COMPLIANCE WITH ETHICAL STANDARDS

This research study was conducted retrospectively using human subject data made available in open access by the references[13]. Ethical approval was not required as confirmed by the license attached with the open access data.

## 6. ACKNOWLEDGMENTS

This work was supported by the National Natural Science Foundation of China (Grant No. 31900979, U1811461) and Zhengzhou Collaborative Innovation Major Project (20XTZX11020).